\documentclass[11pt]{article}

\usepackage[margin=1in]{geometry}
\usepackage[utf8]{inputenc}
\usepackage[T1]{fontenc}
\usepackage{hyperref}
\usepackage{url}
\usepackage{booktabs}
\usepackage{amsfonts}
\usepackage{amsmath}
\usepackage{nicefrac}
\usepackage{microtype}
\usepackage{xcolor}
\usepackage{graphicx}
\usepackage{enumitem}

\title{Evidence-Governed Deployment Readiness for\\
       Networked Agents on Industrial OT Infrastructure}

\date{}

\author{%
  Dibyajyoti Nath \quad Nitya Sathyavageeswaran \quad Ziran Min \quad Pavan Athreya Narasimha Murthy
}

\begin{document}
\maketitle

\begin{abstract}
Deploying a workload on a cyber-secure industrial operational-technology (OT) network requires many cross-layer prerequisites to hold at once, including persistent addressing, name resolution, time synchronization, certificate trust, proxy traversal, and reachable industrial-protocol endpoints. These are owned by different teams, yet current tools treat pre-deployment checking, fault attribution, remediation, and ongoing compliance as separate problems, so evidence is rarely reused across stages. As edge platforms come to host self-adapting agents on shared infrastructure, this fragmentation becomes a governance problem. All four stages ask one question: do a workload’s requirements hold on a given network?

We present an evidence-governed framework that answers it through a single reference, a versioned network-intent artifact specifying a workload’s cross-layer requirements for a specific release. A coverage-driven test loop attaches tested evidence to each requirement before deployment. At a site, readiness is reported as \emph{ready}, \emph{not ready}, or \emph{undecidable}, separating untested requirements from genuine failures; a repair step proposes fixes that preserve co-located workloads; and a drift-watch component re-evaluates the site on change, extending assurance across the workload’s lifetime. A multi-workload OPC~UA commissioning case illustrates all four stages. By detecting runtime deviations and suspending an agent’s guarantees until a refreshed artifact is registered, the framework grounds auditable governance for adapting agent collectives on shared industrial infrastructure.
\end{abstract}

\section{Introduction}
\label{sec:intro}

Industrial edge platforms increasingly host a mix of virtual programmable logic
controllers, SCADA and HMI applications, edge-analytics services, and cloud-connected
gateways on one cyber-secure operational-technology (OT) network.  Each workload arrives
with a precise set of network prerequisites, including a site-approved persistent
address, functioning name resolution, sub-millisecond time synchronisation, a trusted
certificate chain, a valid proxy configuration, correct access-control rules, and
reachable industrial-protocol endpoints.  These prerequisites are owned by different
groups: the network-operations team controls addressing and reachability, the
IT-security team controls certificates and access rules, and the application team
controls endpoint behaviour.
No single group holds a complete view, so a workload can be commissioned only when
all three coordinate, and a requirement that any one of them overlooks surfaces as a
deployment failure.  These workloads are increasingly autonomous and self-adapting,
and a plant hosts a continually adapting collective of them, so the same
coordination gap that complicates a one-off deployment also governs capability
discovery, safe modification, and lifecycle governance of the whole collective on
shared infrastructure over time.

Despite long-standing work on intent-based networking, application descriptors, and
network verification, commissioning a workload in these settings is still largely a
manual, trial-and-error task, as illustrated by the multi-workload OPC~UA
commissioning case we describe in Section~\ref{sec:discussion}.  We attribute this to
five well-defined problems that recur across deployments.
\textbf{(P1)}~No analytical pre-deployment check exists, so whether a site satisfies
every mandatory requirement of a workload can be learned only by deploying the workload
and watching how it behaves.
\textbf{(P2)}~Fault attribution stays manual, because no mechanism traces a failure
back to the requirement it violates, forcing engineers to reconcile documentation,
configuration snapshots, and diagnostic logs across the network, security, and
application layers by hand.
\textbf{(P3)}~Remediation is blind to neighbouring workloads, since existing repair
tools act on network state alone, never model the workload side, and cannot judge
whether a workload-side change would be less disruptive than a network-side one.
As a consequence, a fix that resolves one workload may silently invalidate a
requirement of a co-located one.
\textbf{(P4)}~Post-deployment hardening can break live workloads without warning, as
tightening a firewall rule, narrowing an access-control list, or retiring a DNS record
can invalidate a dependency that happened to be idle when the change was made.
The failure then only surfaces when that dependency is next exercised, possibly during
a safety-critical operation.
\textbf{(P5)}~Continuous compliance checking is impractical, because full-site packet
capture raises privacy concerns and does not scale, and many sites forbid active
enumeration for fear of disturbing safety-critical devices, so a site that was
compliant at commissioning time drifts out of compliance without notice and the
deviation surfaces only when a workload eventually fails in production.

\medskip
The five problems share one root cause.  No shared, application-owned artifact
records \emph{tested} behaviour against a specific workload version and is
consulted by every deployment stage as a common reference.  Our contributions
supply that artifact and the machinery built on it:
\begin{itemize}[leftmargin=1.2em,itemsep=1pt,topsep=2pt]
  \item a versioned network-intent artifact whose clauses carry tested,
        version-bound evidence statuses;
  \item an offline coverage-driven qualification loop that produces those
        statuses without touching live customer devices;
  \item a three-outcome readiness check (\emph{ready}, \emph{not ready},
        \emph{undecidable}) that separates evidence gaps from genuine
        failures, together with a cross-workload constrained repair
        procedure and a passive drift monitor that keep the check current
        over the workload's operational lifetime.
\end{itemize}

\section{Related Work}
\label{sec:related}

Prior work on the four deployment stages falls into three areas, each of which
stops short of a shared, evidence-bearing reference.
\textbf{Application and service descriptors} such as ETSI MEC
\cite{etsi_mec_v1,etsi_mec_v4}, OASIS TOSCA \cite{tosca_yaml,tosca_v2}, and
OPC~UA FX Parts~81 and~82 \cite{opcua_fx81,opcua_fx82} state \emph{what} a
workload needs from its environment, yet none records \emph{whether those needs
have been tested} on a concrete network, and none supports a site-specific
compatibility assessment before deployment.
\textbf{Intent-based networking and orchestration}, from the RFC~9315
\cite{rfc9315} lifecycle and its service-model refinements
\cite{rfc8309,rfc8969,rfc9417} to industrial patents
\cite{us20220124009,us20240259465,us11943101,us10747521} and academic systems
\cite{adn_hotnets23,accord_ucc21,hints_iot23}, makes intent programmable, but
the intent itself is asserted rather than qualified: an operator can declare
that a service requires low latency, yet no artifact records whether that
latency has been measured for this workload version on this network, and no
readiness outcome distinguishes an untested clause from a violated one.
\textbf{Network verification, synthesis, and repair} tools, including MUD
\cite{rfc8520}, Batfish \cite{batfish}, Minesweeper \cite{minesweeper}, SyNET
\cite{synet}, NetComplete \cite{netcomplete}, Aura \cite{aura}, and
edge-oriented variants \cite{ibn_iot_edge}, reason rigorously over network
state, but only over network state: none ingests application-owned evidence
qualified against a specific workload version, and none checks that a repair
respects the requirements of co-located workloads.
The framework in Section~\ref{sec:framework} supplies the missing artifact and
thereby closes the three gaps at once: an analytical readiness check where
descriptors leave an evidence gap, cross-workload repair where verification
tools have a workload-side blind spot, and continuous drift monitoring where
intent-based systems leave compliance unguarded.

\section{Framework}
\label{sec:framework}

\begin{figure}
  \centering
  \includegraphics[width=0.5\linewidth]{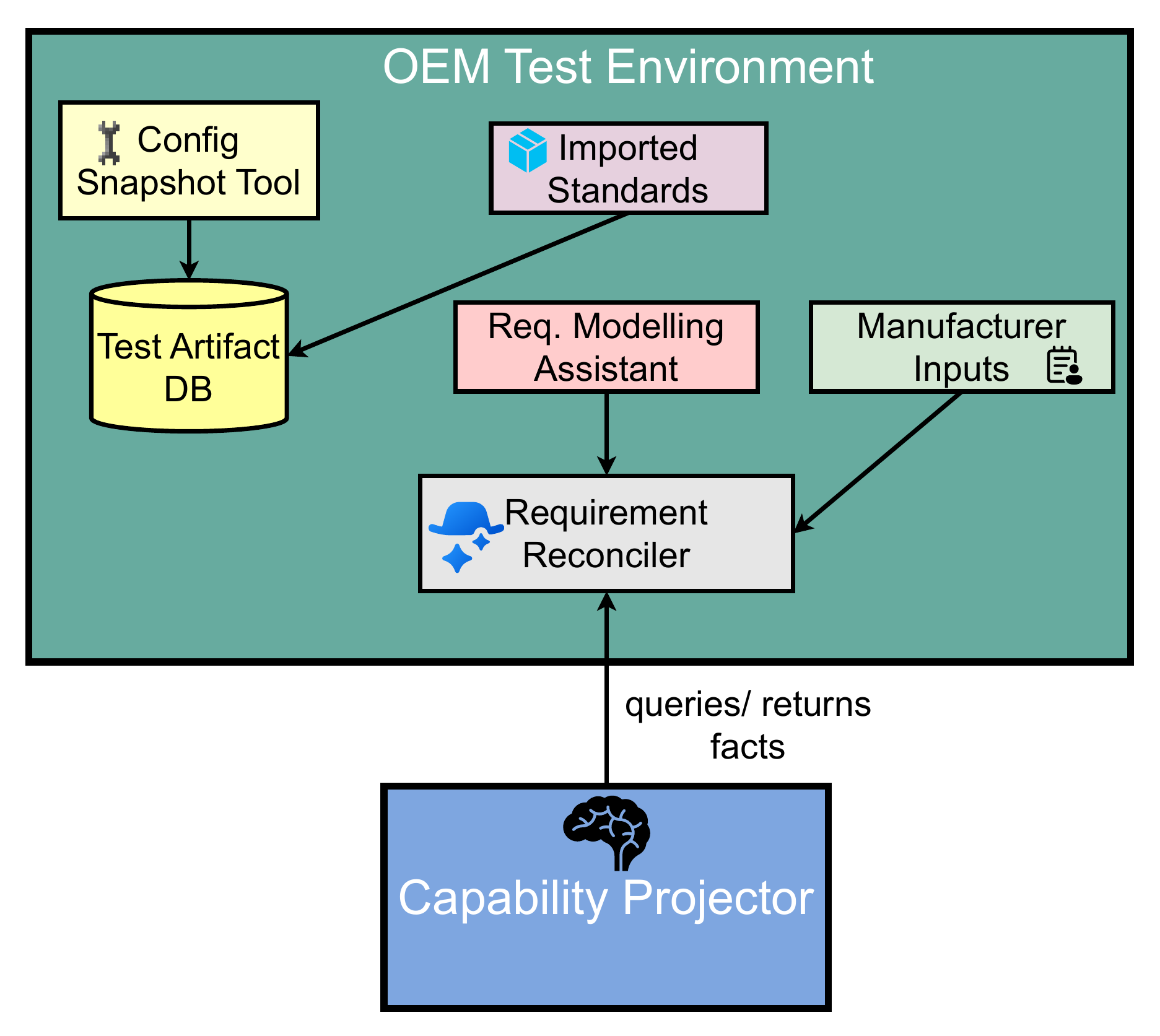}
  \caption{Artifact creation and evidence qualification cycle in the test
           environment (referenced from Section~\ref{sec:framework},
           Stage~1).  The Capability Projector constructs test configurations
           clause by clause; the Requirement Reconciler records version-bound
           pass or fail outcomes until the coverage threshold $\tau$ is
           reached.}
  \label{fig:testbed}
\end{figure}

\begin{figure}
  \centering
  \includegraphics[width=0.5\linewidth]{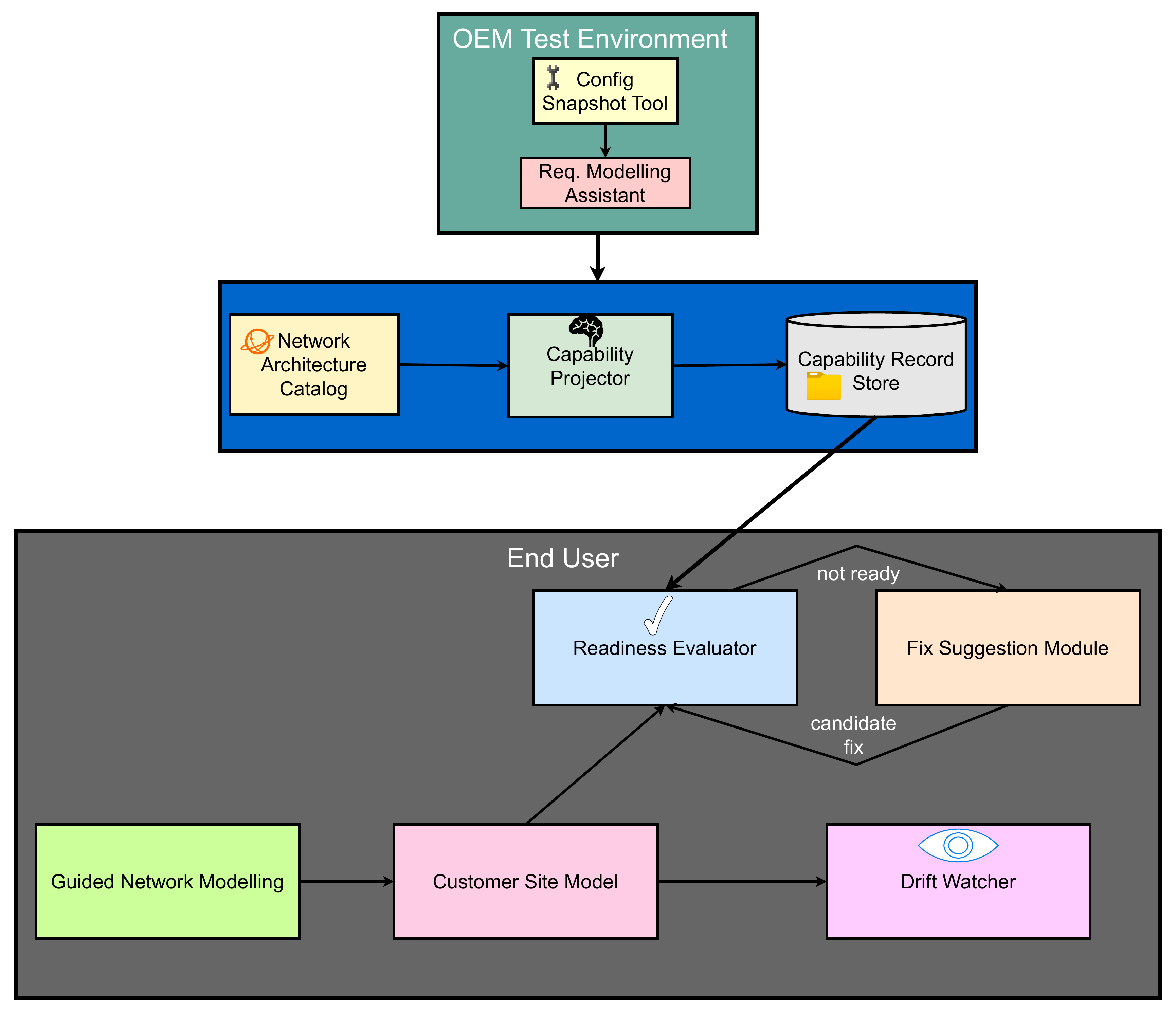}
  \caption{Multi-agent architecture for artifact-driven deployment at the
           end user (referenced from Section~\ref{sec:framework},
           Stage~3).  The Readiness Evaluator, Repair Planner, and Drift
           Watcher operate continuously against the End User Model and
           the shared evidence database.}
  \label{fig:site}
\end{figure}

Sections~\ref{sec:intro} and~\ref{sec:related} established that the four
deployment stages should share a single artifact.  This section defines that
artifact and the four stages that operate on it.  The stages are carried out
by five cooperating agents that read from and write to a shared evidence
database and a shared site model: the \emph{Capability Projector} and
\emph{Requirement Reconciler} run offline in Stage~1, while the
\emph{Readiness Evaluator}, \emph{Repair Planner}, and \emph{Drift Watcher}
run on-site in Stages~2 and~3 and in the drift loop.  Figures in
Appendix~\ref{app:figures} illustrate the offline evidence-qualification loop
(Figure~\ref{fig:testbed}) and the on-site multi-agent architecture
(Figure~\ref{fig:site}).

\paragraph{Network-intent artifact.}
The central object of the framework is a \emph{versioned network-intent
artifact} bound to a specific workload version.  It records every network and
application-configuration requirement that the workload needs for correct
operation as a set of \emph{requirement clauses}, each of the granularity of a
single verifiable statement, for example ``DNS resolves
\texttt{historian.plant.local} to a reachable address'' or ``UDP port 4840 is
accessible on all configured OPC~UA server endpoints''.  Every clause is paired
with an \emph{evidence status} drawn from
\begin{equation}
  \mathcal{S} = \{\,\textit{demonstrated},\;\textit{partially supported},\;
  \textit{unsupported},\;\textit{contradicted}\,\},
  \label{eq:evidence-status}
\end{equation}
and tagged as \emph{mandatory} (its status gates readiness) or \emph{optional}
(status is recorded but does not block deployment).  Alongside the clauses, the
artifact records workload identity and version, role relationships, service
dependencies, addressing and name-resolution behaviour, time synchronisation,
multicast membership, path reachability, policy handling, and operational safety
bounds.  The artifact is fixed at workload-version creation time and is never
modified by site-specific evaluation, so every subsequent stage reasons against
a stable object.

\paragraph{Stage 1: evidence qualification.}
Evidence qualification is the offline stage that turns bare clauses into
statuses in $\mathcal{S}$ (Figure~\ref{fig:testbed}).  Let
$\kappa(c)\in[0,1]$ denote the fraction of the enumerable distinguishable test
configurations for clause $c$ that have already been executed and recorded, and
let $\tau$ be a configurable threshold on that fraction.  The test space is
bounded per clause by the finite set of variables that clause names (for
example, address family, DNS record type, or TLS-version and cipher choice),
so enumeration walks a small combinatorial product, not the site-wide
configuration space; unnamed variables are held to representative
defaults, and $\tau$ trades assurance against qualification cost per clause.  For every clause with
$\kappa(c)<\tau$, a \emph{Capability Projector} constructs a minimal network
configuration in a controlled test environment designed to exercise $c$; a
\emph{Requirement Reconciler} evaluates the resulting state against $c$ and
appends an outcome
$(c,v_\mathrm{wl},v_\mathrm{evidence},r)$, where $v_\mathrm{wl}$ is the workload
version to which the artifact is bound, $v_\mathrm{evidence}$ is a monotonic
counter of the evidence database at the time of the test, and
$r\in\{\text{pass},\text{fail}\}$, and updating $\kappa(c)$.  Coverage is tracked
per clause, and $\tau$ may be set higher for mandatory clauses than for optional
ones; a variable shared across several clauses (for example, TLS version) is
enumerated once but counted towards each clause it appears in.  The loop terminates
when $\kappa(c)\geq\tau$ holds for every clause or no further distinguishable
test conditions can be constructed.  The database is the sole source of clause
status values: a clause becomes \emph{demonstrated} only after every recorded
outcome is a pass and $\kappa(c)\geq\tau$, \emph{contradicted} as soon as one
recorded outcome is a fail, \emph{partially supported} when it has at least one
pass, no fail, but $\kappa(c)<\tau$, and \emph{unsupported} when it has never
been exercised.  This last case is preserved through all downstream reasoning
and is the basis of the \emph{undecidable} verdict.

\paragraph{Stage 2: site readiness assessment.}
Deploying the workload at the end user raises the analytical question that
P1 identifies as unanswered today: does the site satisfy every mandatory clause
of the artifact?  Let $A$ denote the artifact and $M$ a \emph{Customer Site
Model}, a structured snapshot of the site's configuration, reachability, naming
and policy state assembled from customer-controlled sources such as
configuration-management exports, IPAM and DNS records, firewall and
access-control policies, and PKI trust stores.  Every verdict is conditioned on
the fidelity of $M$: an omission or stale entry can turn a genuine mismatch
into a spurious \emph{ready}, so $M$ is versioned alongside the evidence records
that consult it, and clauses whose declared value in $M$ disagrees across two or
more independent sources (for example, IPAM and DNS for the same host) are
flagged as \emph{unsupported} rather than accepted, which routes them to
\emph{undecidable} and prevents undetected drift in $M$ itself from silently
producing a false \emph{ready}.  Given $A$ and $M$, the \emph{Readiness Evaluator} checks each
mandatory clause $c\in A$ against $M$ and returns \emph{ready} if every such
clause is satisfied in $M$ and backed by at least one demonstrated or partially
supported pass record, \emph{not ready} if any mandatory clause is directly
violated by $M$, and \emph{undecidable} if no clause is violated but at least
one lacks any applicable evidence record.  The three-valued outcome matters
because it treats an evidence gap and a genuine failure differently: the
first is logged with its unresolved clauses and triggers targeted evidence
collection, whereas the second is annotated with the clauses and records
that caused it and returned to the operator for repair.

\paragraph{Stage 3: cross-workload constrained repair.}
A \emph{not ready} verdict alone does not fix the site.  The \emph{Repair
Planner} agent (Figure~\ref{fig:site}) generates candidate configuration
changes $\Delta$ against $M$, where $M\oplus\Delta$ denotes the site model
obtained by applying $\Delta$ to $M$ with entries in $\Delta$ overriding the
corresponding entries in $M$.  The Repair Planner does not act on the site
directly; it queries the Readiness Evaluator for each candidate and only the
authorised $\Delta$ is handed to the operator or to a downstream
configuration-management agent for application.  Each candidate is subjected
to a three-step check: (i)~re-run
the readiness check for the failing artifact under $M\oplus\Delta$; (ii)~re-run
it for every other active workload artifact bound to the same site;
(iii)~authorise $\Delta$ only if step~(i) succeeds and step~(ii) produces no
regressions.  A fix that would silently invalidate a co-located clause is
therefore rejected before it is applied.  Among authorised candidates,
workload-side changes such as endpoint reconfiguration or proxy injection are
preferred to network-wide changes such as ACL expansion, because the former
affect fewer co-located workloads.  Two structural constraints keep the
candidate space tractable: candidates are generated only for the clauses
that Stage~2 marked \emph{not ready}, and each candidate is confined to
the configuration scope of the clause it addresses (for example, a
DNS-record change for a naming clause, or a rule insertion into the ACL
that governs the failing path).  The branching factor
of the search thus scales with the number of failing clauses, not with the
site.  When the candidate space is exhausted without a \emph{ready} outcome,
the site is reported as \emph{unrepairable within the current search space}
and escalated together with the enumerated candidates and their rejection
reasons; a genuine absence of an admissible fix is then distinguishable from
a search-space limitation.

\paragraph{Continuous drift monitoring.}
Readiness at commissioning time does not imply readiness a month later.  A
\emph{Drift Watcher} maintains $M$ from operator inputs and passive
observations (configuration-management exports, syslog events, SNMP traps),
without active probing of customer devices.  Any state change $\delta M$
triggers an incremental re-evaluation of the clauses it may affect, and a
full readiness pass follows if any status changes, so assurance extends
across the operational lifetime of every workload without the packet capture
or active enumeration that make P5 impractical.  Passive observation has a
known limit: changes that leave no such record, such as a silently dropped
firewall rule, remain invisible until they cause a downstream event.  A
scheduled full snapshot of $M$ diffed against the incremental $\delta M$
stream bounds this window in practice.

\section{Discussion and Conclusion}
\label{sec:discussion}

\paragraph{Resolution of P1--P5.}
The four stages of Section~\ref{sec:framework} map onto the five problems of Section~\ref{sec:intro} in a direct correspondence.  Stage~2 answers P1 by deciding readiness against the artifact before deployment, and P2 because every \emph{not ready} verdict is annotated with the clauses and records that produced it. Stage~3 answers P3 by admitting workload-side candidates
alongside network-side ones and rejecting any $\Delta$ that regresses a co-located artifact.  The Drift Watcher answers P4 and P5 through passive re-evaluation of $M$, without the packet capture or active probing that make continuous compliance impractical. Two properties set the framework apart from prior descriptors and repair tools: each clause carries a tested status, so \emph{tested-and-failed} is distinguishable from \emph{never-tested}; and the \emph{undecidable} verdict keeps evidence gaps from being reported and remediated as failures.

\paragraph{Illustrative case.}
Consider three workloads on one OT segment: an OPC~UA historian, a SCADA client, and an edge-analytics service, each shipping an artifact with clauses on naming, PTP time synchronisation, certificate trust, and TCP~4840 reachability.  Suppose PKI trust is \emph{undecidable} for the analytics service because the test environment did not include the site's CA chain, so the corresponding clause remains \emph{unsupported} rather than
contradicted; the SCADA client, by contrast, is \emph{not ready} on TCP~4840 because $M$ shows the port blocked by an existing ACL. Stage~3 proposes a workload-side proxy for the analytics service and a scoped ACL insertion for the SCADA path; both are checked against the historian's clauses before authorization, so unblocking the SCADA client cannot silently break the historian. After commissioning, suppose the analytics service opens a new outbound connection to a cloud endpoint not listed in its mandatory clauses: the hook event is flagged as an \emph{artifact--behavior disagreement}, and the service's Stage~3 authorizations are suspended until a refreshed artifact is registered.

\paragraph{Limitations and outlook.}
Readiness quality is bounded by how well the test environment mirrors real end users, and the per-clause repair search becomes expensive on clauses defined over large ACL spaces.  The local-hook mechanism has three open boundaries: it trusts participating agents to emit accurate events,
so a compromised one that forges or suppresses events needs complementary detection; it requires the workload owner to be authorized to install the hook, which in OT settings with distinct site and workload operators is a
contractual question; and it does not cover legacy or third-party
components outside the framework's perimeter.  Convergence guarantees for the repair cycle, federated evidence pooling, online clause learning for
behavior-only adaptation, hook-independent attestation, and a shared OT benchmark are left to future work.  What the framework already delivers is a single auditable lifecycle in place of four disjoint manual checks: one versioned, evidence-bearing artifact, consulted by every stage, that grounds capability representation and governance for continually adapting agent collectives on shared industrial infrastructure.

\bibliographystyle{unsrt}
\bibliography{references}

\newpage
\appendix
\section*{Appendix: Figures}
\label{app:figures}
\addcontentsline{toc}{section}{Appendix: Figures}

\end{document}